# Direct Observation of Re-entrant Multiferroic CuO at High Pressures

Rajesh Jana[1], Pinku Saha[1], Vivek Pareek[1], Abhisek Basu[1], Guruprasad Mandal[1], Sutanu Kapri[2], Sayan Bhattacharya[2] and Goutam Dev Mukherjee[1*]

[1]Department of Physical Sciences, [2]Department of Chemical Sciences, Indian Institute of Science Education and Research Kolkata, Mohanpur Campus, Nadia – 741246, West Bengal, India.

## ABSTRACT

We have carried out a detailed experimental investigation on CuO using dielectric constant, ac resistance, Raman spectroscopy and X-ray diffraction measurements at high pressures and room temperature. Both dielectric constant and dielectric loss show anomalous peak in the pressure range 3.4 – 4.4 GPa indicating a ferroelectric transition. Raman studies show anomalous behaviour of the $A_g$ mode with a slope change in the mode frequency and a minimum in the mode FWHM at 3.4 GPa indicating a strong spin phonon coupling along [1 0 -1] direction. A step like behaviour in the intensity of the $A_g$ is observed at 3.4 GPa, indicating a change in polarization of the mode. A maximum in the intensity of the (2 0 -2) Bragg peak at 3.4 GPa points to the occurrence of critical scattering due to emergence of magnetic exchange interaction. All our experimental evidences show to the presence of re-entrant type-II multiferroic behaviour in CuO at 4 GPa.

*Corresponding Author

**Introduction**

In recent years cupric oxide (CuO) has generated a renewed interest in the scientific community as it holds a promise to be the room temperature type II multiferroic with large polarization. Polycrystalline CuO is a p-type semiconducting material with a narrow band gap of 1.2 eV [1]. It has attracted special interest due to discovery of high $T_C$ superconductivity in cuprates and other wide applications in the industrial field, such as, fabrication of solar cells [1-2]; lithium ion batteries [3]; magnetic storage media, gas sensors [4] etc. CuO is found to be quasi one dimensional (1D) antiferromagnet with a high Neel temperature ($T_N$) of 230 K due to its large antiferromagnetic exchange interaction along [1 0 -1] direction [5-7]. Kimura *et al.* [8] showed that in the temperature range, 213K< T<230K it exists in an in-commensurate antiferromagnetic phase, which is ferroelectric with the Curie temperature, $T_C$ = 230K and hence behaves as a type II multiferroic in this short temperature range. High pressure neutron diffraction studies up to 1.8 GPa showed that $T_N$ increases to 235 K [9]. Recent theoretical studies have shown that the antiferromagnetic super exchange parameter in CuO increases with pressure and in the pressure range 20-40 GPa it is predicted to exist in the multiferroic state at room temperature [10]. High pressure X-ray diffraction and Raman spectroscopy measurements up to about 47 GPa have shown nanocrystalline CuO remains stable in its monoclinic (space group C2/c) state up to about 47 GPa [11].

In the present work we propose that multiferroic phase may be induced in bulk CuO in the pressure range about 3.4 - 4 GPa at room temperature, much earlier than predicted from theoretical studies. Our conclusion is based on detailed high pressure experiments from dielectric constant, ac electrical resistance measurements; Raman spectroscopy and X-ray diffraction studies.

**Experimental**

CuO was synthesized in laboratory using a typical synthesis protocol where two aqueous solutions were prepared with distilled water: 4 g of $CuCl_2.2H_2O$ in 100 ml $H_2O$ and 2 g of NaOH in 30 ml $H_2O$. The NaOH solution was added slowly into $CuCl_2.2H_2O$ solution under vigorous stirring. The hydrolyzed product was centrifuged followed by washing with distilled water. The precipitate from the centrifuge tube was dried in air at 80 $^o$C for 3 hrs and further heat treated in an alumina boat in air at 800 $^o$C for 1 h. Upon cooling the furnace to room temperature, the black coloured product was ground in a mortar pestle for further use. The above washing and drying procedure was repeated few times to improve the purity of the CuO powder. The final product was characterized for phase purity by X-ray diffraction measurements using the 9 kW rotating anode high resolution Rigaku SmartLab X-ray diffractometer with Cu $K_\alpha$ X-ray source. All the observed diffraction peaks are indexed to monoclinic structure with space group C2/c and unit cell parameters: $a_0$ = 4.6946(3), $b_0$ = 3.4182 (2), $c_0$ = 5.1329 (3), $V_0$ = 81.202(7), $\beta_0$ = 99.65(4), which are in good agreement with the literature [11- 12 ].

High pressure dielectric constant and dielectric loss and electrical resistance measurements were carried out with a GW Instek make Precision LCR Meter, Model-800G. We used a Toroid anvil (TA) apparatus along with hydraulic press for the above measurements for pressures up to about 9 GPa. The TA apparatus was pressure calibrated using Bi I-II and Yb hcp-bcc transitions at 2.65 GPa and 4 GPa respectively. A pellet of the sample of 3 mm diameter and 1 mm thickness was sandwiched between two copper plates of thickness 0.1 mm and placed inside a Teflon cup with a lid. Teflon cup was inserted into the hole of diameter 5 mm drilled at the central part of the toroid shaped pyrophylite gasket. Two thin steel wires of 20 micron dia were attached to the copper plate and taken out through small holes in the Teflon cup and then through the side holes of the gasket. The gasket with whole sample assembly was compressed and locked between two toroid shaped opposed anvils by the 300 ton hydraulic press for 30 minutes at initial pressure 0.5 GPa to ensure a good contact between electrodes and the sample. The initial compression reduced the sample thickness to about 0.5 mm and no further reduction has been observed with increasing pressure up to 9 GPa and back to ambient condition. Also no deformation has been observed in the Teflon cup, which ensures a reproducible data in the pressure range of 0 - 9 GPa. AC measurements were carried out in the frequency range 33 Hz -1 MHz. Initially a contact resistance of 230 Ω was measured mainly coming from the long 20 micron steel wire. Short circuit trimming of LCR Meter was done to calibrate the contact probes for stray capacitance and series impedance of different metal contacts and steel wire. At each pressure 20 successive data were taken for the precision in measurements.

Pressure dependent dielectric constant was calculated from the measured capacitance using the relation, $C(P) = \varepsilon(P) \varepsilon(0) (A(P)/d(P))$, where $\varepsilon(P)$, $A(P)$ and $d(P)$ are dielectric constant, area of the copper plates and thickness of the sample respectively at pressure P and $\varepsilon(0)$ is the free space permittivity. Area of the copper plates can be assumed to be constant for pressures up to 9 GPa

Raman spectroscopic measurements at high pressures and room temperature were carried out in our laboratory using Horiba Jobin Yvon LabRam HR 800 Raman spectrometer equipped with an air-cooled charge coupled device detector. High pressure was generated using a diamond anvil cell (Easylab Co.,UK) with diamond culets of size 300 μm. For Raman measurements sample was loaded together with KCl as pressure transmitting medium, into the sample chamber of 100 μm hole in a preindented stainless steel gasket of thickness of about 45 μm. Few ruby chips of size 2-3 micron were also loaded along with sample and pressure was measured using ruby fluorescence technique [13]. The sample was excited with the $Ar^+$ ion (488 nm) laser of power 6 mW and Raman signal was collected in back scattering geometry with an 1800 g/mm grating with a resolution of 1.2cm$^{-1}$. A long working distance infinitely corrected 20X objective was used for both focussing of laser and collection of Raman signal.

High pressure x-ray diffraction measurements were performed at room temperature at XRD1 beamline in Elettra synchrotron source, Trieste, Italy. The sample along with silver powder as pressure marker were loaded in the 100 μm hole of a preindented steel gasket of thickness 45 μm and a 4:1 methanol-ethanol mixture was used as pressure transmitting medium. Pressure was determined by the equation of state (EoS) of silver [14]. X–Ray patterns were collected employing a monochromatic X-Ray beam of wavelength 0.689 A, on a 2M Piltus (Dectris, Switzerland) Si pixel detector. The X-Ray was collimated to 70 μm diameter. Two dimensional diffraction images were integrated into intensity versus 2θ profile using FIT2D software [15]. The collected XRD patterns were indexed using freely available software Dicvol [16] and then analyzed using the Rietveld refinement program GSAS [17].

**Results and Discussion**

We have carried out frequency dependent dielectric constant and dielectric loss measurements using parallel plate capacitor method up to about 9 GPa. At very low pressures, dielectric constant is estimated to be 3.75 x 10$^3$ at a frequency of 1 KHz. Such high value of dielectric constant can be attributed to the presence of small amount of Cu$^{3+}$ ions in the powder sample [18-19]. Frequency dependent dielectric constant and dielectric loss measurements show large dispersion in the data indicating hopping of holes between Cu$^{2+}$ and Cu$^{3+}$ ions [19]. Fig.1 shows the pressure dependence of dielectric constant and dielectric loss, at a frequency of 1 KHz. As pressure increases, ε(P) remains almost constant up to about 2 GPa and then rapidly rises up to 4 GPa, followed by a sharp drop. Dielectric loss behaviour shows an anomalous peak in the range 3.7 to 4.7 GPa. Both these sharp anomalies in ε(P) as well as in dielectric loss are indicative of ferroelectric transition. The frequency dependence of dielectric constant at various pressures shows an interesting behaviour (Fig. 2). In all cases dielectric constant reduces with frequency showing normal dispersion. At low frequencies, ε(P), increases rapidly up to about 3.4 GPa and then above 4 GPa they all lie on the same curve. This may be due to the fact that in the ferroelectric region, above 4 GPa, all the dipoles are aligned and hence unless the applied electric field is too strong the orientational polarizability of dipoles will not contribute.

The variation of relative change in ac resistance (R(P)) with respect to the resistance measured at 0.17 GPa (R(0.17)) with pressure shown in Fig.3. The resistance drops by almost three orders of magnitude in the range about 3 – 4.5 GPa. Similar instabilities in resistance values has been observed in sintered semiconducting CuO pellets with a resistance drop more than two orders of magnitude below 230 K, that coincides with the antiferromagnetic transition in the sample [20]. Electron diffraction experiments at ambient temperature has shown the presence of quasi 1D zigzag charge stripes in p-type CuO [21] indicating the presence of strong spin-charge coupling in the system. Sudden decrease in resistance in CuO above 3 GPa indicates that onset of change in polarization in the sample facilitates movement of trapped charges under the application of external electric field. In polycrystalline CuO powder the conduction mechanism is governed by holes due to presence of $Cu^{3+}$ ions. Hopping of holes produces $Cu^{3+}$ ions around antiferromagnetically ordered $Cu^{2+}$ ions, and in turn interacts with the spins. Above 4 GPa, complete onset of long-range antiferromagnetic order can pin the charges and stop their movement. This probably results in sudden increase of resistance at 4 GPa.

We have carried out high pressure Raman spectroscopic studies of polycrystalline CuO up to 42 GPa. Selective Raman spectra at different pressures are shown in Fig. 4. Since, there are two molecular units in the primitive cell of monoclinic CuO, from the group analysis [22] decomposition of normal modes of vibration at zone centre is given by,

$$\Gamma = 4A_u + 5B_u + A_g + 2B_g$$

One finds 12 normal modes in total, six IR active modes ($3A_u + 3B_u$), three acoustics modes ($A_u + 2B_u$) and only three ($A_g + 2B_g$) modes are Raman active [22]. It can be seen from Fig. 4 that our ambient Raman spectrum contains three peaks at 292, 342 and 625 cm$^{-1}$ in close agreement with earlier reported values [23, 24]. The most intense peak at 292 cm$^{-1}$ is assigned to $A_g$ mode and other two less intense peaks are assigned to $B_g$ modes [24]. With increasing pressure all three peaks linearly shift towards the higher frequency region and become broader. Absence of emergence of any new Raman mode indicates to no structural changes within the pressure range of this study. On complete release of pressure three Raman modes match with that of parent material showing the reversibility of pressure effect.

The Raman spectra were normalized with respect to the Bose Einstein thermal factor by dividing the raw spectra by the factor ($n(\omega)+1$), where $n(\omega) = 1/(\exp(-E/k_BT)-1)$; E is the energy of the mode, $k_B$ is the Boltzman constant and T is the room temperature value. All the modes are fitted with the standard Lorentzian function. The different fitting parameters of the $A_g$ mode show several interesting anomalous changes. The frequency of $A_g$ mode increases linearly with pressure, however there is a definite change in slope at about 3.4 GPa (Fig. 5a). The slope decreases from 4.8(4) cm$^{-1}$/GPa below 3 GPa to 2.4(1) cm$^{-1}$/GPa above 3 GPa. In the absence of any structural change at 3 GPa such a small change of slope in Ag mode can be attributed to small change in Grüneisen parameter arising from electronic contribution. The full width half maximum (FWHM) of the $A_g$ mode decreases rapidly till about 3.4 GPa and then increases with pressure (Fig. 5b). The FWHM of a Raman mode is related to the lifetime of the phonon and it may get affected due to coupling of phonons to electrons or their spins. A minimum is indicative of a phase change. Since there is no indication of a structural transition, the observed minimum in the FWHM of the Ag mode at about 3.4 GPa can only be related to an electronic transition. As CuO is an insulator type, one can attribute the minimum in FWHM of the Raman mode to a spin-phonon coupling process. This assumes significance in view of the fact that Ag mode is associated with the oxygen vibrational mode parallel to the b crystal axis, i.e., the monoclinic axis, and the multiferroic behaviour of CuO has been observed due to stronger antiferromagnetic exchange interaction along [10-1] direction. One can relate the above anomalous changes in the pressure evolution of the frequency of Ag mode and it's FWHM at about 3.4 GPa to spin-phonon interaction due to re-entrant antiferromagnetism in CuO.

In Fig. 5c we have plotted the normalized intensity of the $A_g$ mode with respect to pressure, which shows a step like behaviour with an anomalous jump at 3.4 GPa. Raman scattering intensity is directly proportional to the square of mode polarizability. Therefore the sudden increase in the intensity can be attributed to change in polarization of the Cu-O-Cu bonding line due to strong dynamic O-ion displacements.

To understand pressure induced effect of CuO in more detail we have conducted careful *in situ* XRD measurements under pressure as a complementary study to Raman measurements. The ambient and a few selective angle dispersive XRD patterns with increasing pressure up to 41 GPa are shown in Fig. 6a. Over the whole pressure range, no remarkable changes in the diffraction patterns have been observed except broadening and merging of higher angle peaks. As the pressure increases Bragg peaks shift to higher 2θ values due to lattice contraction and we have not observed any extra peak with pressure that confirms the structural stability of the material, consistent with the earlier studies [11]. Analysis of collected XRD patterns at various pressure were carried out using the program GSAS [17]. For this purpose we have used the structural model of CuO from high pressure neutron diffraction studies [25-26]. The P-V curve (Fig. 6b) for the entire region of pressure has been fitted by a 3$^{rd}$ order Birch-Murnaghan equation of state [27-28],

$$P = (3/2)\ B_0\ [(V_0/V)^{7/3} - (V_0/V)^{5/3}\ ][1 - \tfrac{3}{4}(4 - B') \times \{(V_0/V)^{2/3} - 1\}],$$

where, $B_0$ and $V_0$ are the bulk modulus and volume at ambient pressure respectively. B' is the first derivative of bulk modulus with respect to pressure. Our best fit gives $B_0$ = 104(4) GPa, $V_0$ = 80.85(4), and B' = 5.082(9).

Since we have used liquid pressure transmitting medium and the previous anomalies are observed around 3.4 GPa, in the present case we have restricted analysis of our XRD patterns below 10 GPa. In Fig. 7a we have shown pressure variation of relative change in lattice parameters 'a', 'b', 'c', with respect to the ambient pressure value, which show that they behave in similar manner as observed in earlier neutron diffraction studies. Neutron diffraction experiments have shown that high temperature incommensurate antiferromagnetic state has a wave vector (0.506, 0, -0.483), where as the low temperature antiferromagnetic state locks into a wave vector (0.5,0,-0.5) [9]. Under this context we looked into the pressure variation of background subtracted integrated intensity of (2,0,-2) Bragg peak that is shown in Fig. 7b. Interestingly enough a maximum is observed at about 3.4 GPa and can probably be related to enhanced X-ray scattering due to emergence of antiferromagnetic at the (2,0,-2) Bragg plane. Similar critical scattering effect has been observed by Yang *et al.* [6] at (1/2,0,-1/2) Bragg position from neutron diffraction experiments. Many different studies have shown that Cu-Cu spin exchange parameter increases with increasing Cu-O-Cu bond angle in copper oxide compounds, which may increase the Neel temperature. From pressure dependent neutron diffraction experiments Chatterji *et al.* [9] have shown a increase in Cu-O-Cu bond angle along [10-1] direction and Neel temperature increases to 235 K at 1.8 GPa from a value of 231 K at ambient pressure.

All our experimental evidences from detailed investigations using dielectric constant, resistance, Raman spectroscopy and XRD measurements suggest that polycrystalline CuO may become multiferroic above 4 GPa at room temperature, which is at much lower pressures than that predicted by the theoretical simulations. Several theoretical and experimental studies have shown that the high value of antiferromagnetic superexchange parameter among $Cu^{2+}$ ions along [10-1] direction gives rise to large $T_N$ in CuO. Shimizu *et al.* have proposed that in low dimensional cuprates, antiferromagnetic J value increases with increase in Cu-O-Cu bond angle from 90 – 180 $^0$. This was verified by Rocquefelte *et al.* [10] from their theoretical calculations and they showed that J and Cu-O-Cu bond angle show a systematic variation. It has been shown that increase in J gives rise to magnetic frustration, which breaks the inversion symmetry, and gives rise to multiferroic property in CuO. From neutron diffraction studies Chaterji *et al.* [9] have shown that application of pressure increases Cu-O-Cu bond angle, which results in increase of the value of that translates into increase of $T_N$ of CuO. However the experimental $T_N$ determination was limited to 1.8 GPa. To understand the relation among J, $T_N$ and Cu-O-Cu bond angle of $CuO_4$ tetrahedra, in Fig. 8 we have plotted dependence of J and $T_N$ on the Cu-O-Cu bond angle for various cuprate materials [9, 29-31]. One can see that for certain cuprates even though J value increases with Cu-O-Cu bond angle, $T_N$ drops. Decrease in $T_N$ for $La_2CuO_4$ and $Nd_2CuO_4$ has been attributed to increased magnetic frustration due to presence of anisotropic exchange interaction [29]. This indicates that value of J cannot be the only criteria for deciding $T_N$. At about 4 GPa, Cu-O-Cu bond angle along [10-1] direction in CuO is about 153º, which translates to a value of $T_N$ of about 280 K. This value is estimated from the extrapolation of $T_N$ value with respect to Cu-O-Cu bond angle from CuO to $YBa_2Cu_3O_6$. This is a surprising result that coincides with our experimental investigations considering such a phenomenological model of obtaining $T_N$ from Cu-O-Cu bond angle value in cuprates. In fact Rocquefelte *et al.* [10] have shown that the relation between Cu-O-Cu bond angle and J is non-linear and they have attributed it to the anisotropic change in Cu-O bond lengths. Here we argue that anisotropic compression of the Cu-O bond lengths gives rise a more ordered $CuO_4$ tetrahedra and hence the magnetic disorder increases, that brings down $T_N$ to 3.4- 4 GPa at room temperature. This view is consistent with the spin-phonon coupling in CuO at 3.4 GPa, where the $A_g$ mode shows slope change. $A_g$ mode is attributed to O-atom vibrations with respect to each other and in the simplest case we can model in terms of a 1D diatomic chain of Cu and O atoms. The energy of the phonon at BZ boundary can be given by, $(2K/M_O)^{1/2}$, where K is the force constant and $M_O$ is the mass of the oxygen atom. Small mass of O gives rise to large vibrational amplitudes that can favour large magnetic fluctuations. Since the square of the amplitude of the vibrational mode can be expressed as, $u^2 = 3\hbar\omega/2K$, at the onset of magnetic transition large critical fluctuations will reduce the force constant K. Therefore the phonon frequency seems softer with respect to the paramagnetic phase and hence the slope changes.

**Conclusion**

Our experimental evidences show that CuO can become multiferroic at room temperature at a nominal pressure of 4 GPa. The importance of magnetic disorder due to non-linear behaviour of the lattice is apparent in the realization of room temperature binary type II multiferroic. We feel that our results will induce more research activity for accurate theoretical modelling of the multiferroic systems.

**Acknowledgement**

GDM gratefully acknowledges financial support from Department of Science and Technology (DST), Government of India and Ministry of Earth Sciences, Government of India for financial support for developing the high pressure laboratory for this work. GDM also acknowledges DST, Indo-Italian Program of Co-operation and Elettra Synchrotron Light Source for financial and laboratory support for synchrotron based X-ray diffraction measurements.

Figures with Figure Captions

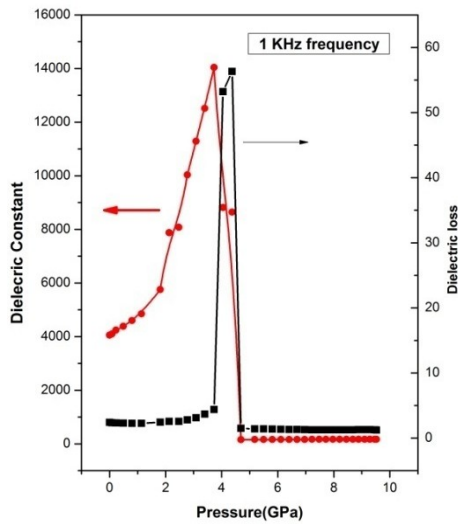

Fig.1: Pressure evolution of dielectric constant (red dots) and dielectric loss (black squares) of CuO.

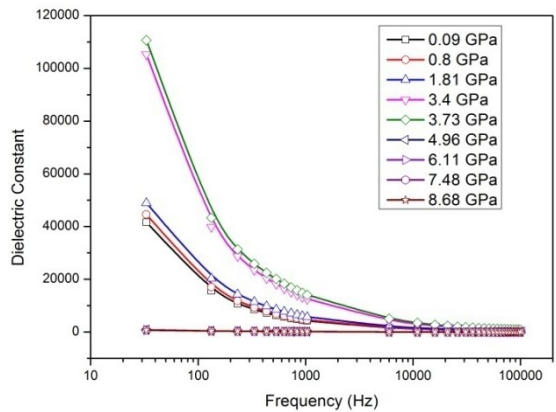

Fig.2: Frequency dependence of dielectric constant at various pressures. The low frequency dielectric constant increases as one approach 3.73 GPa and in the ferroelectric phase dispersion in the data is minimum.

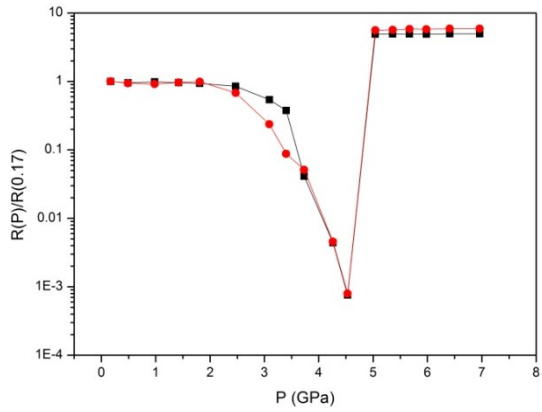

Fig.3: Pressure dependence of ac resistance at 1KHz from two different experiments show a decrease of three orders of magnitude in the pressure range 3 – 4.5 GPa followed by a sudden increase.

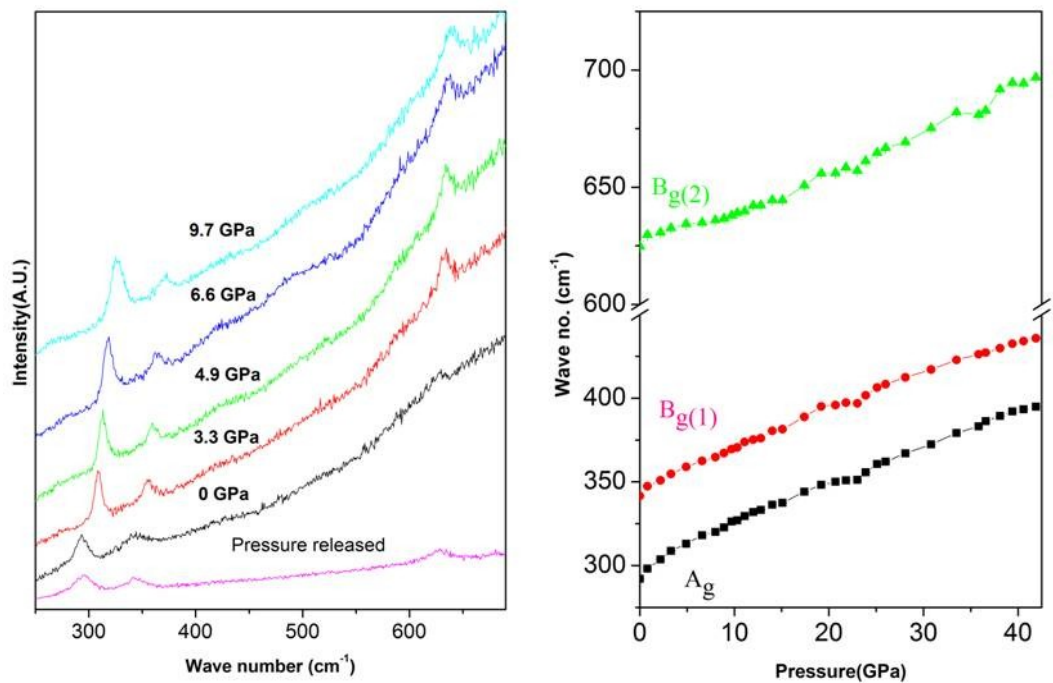

Fig.4: Pressure evolution of (a) Raman Spectra up to about 9.7 GPa and (b) mode frequencies up to about 42 GPa of CuO showing presence of no structural transition.

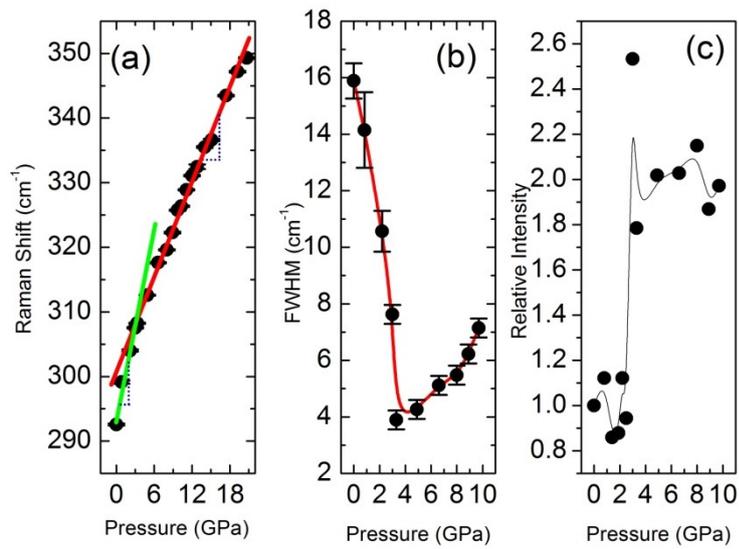

Fig.5: (a) Linear pressure evolution of Ag Raman mode with a slope change at 3.4 GPa; (b) FWHM of Ag Raman mode showing a minimum at 3.4 GPa; and (c) relative intensity of the Ag Raman mode with respect to the ambient pressure value show an abrupt jump at 3.4 GPa indicating a change in polarization of the above said mode.

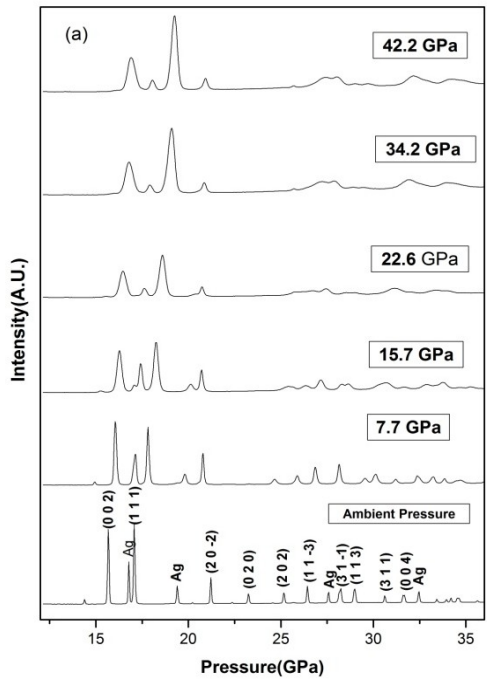
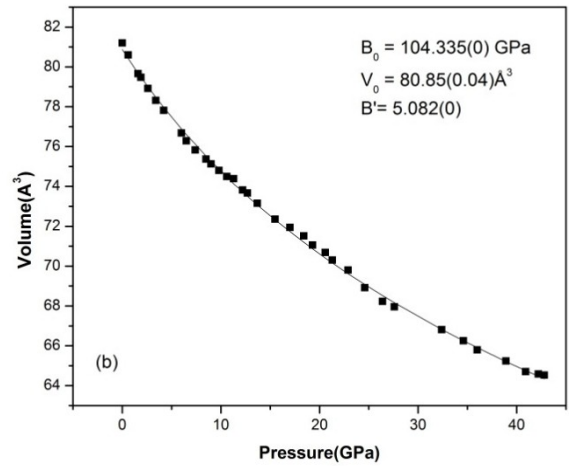

Fig.6: (a) Pressure evolution of X-ray diffraction patterns of CuO till 42 GPa showing no structural transition and (b) Birch-Murnaghan EOS fit to the P~V data.

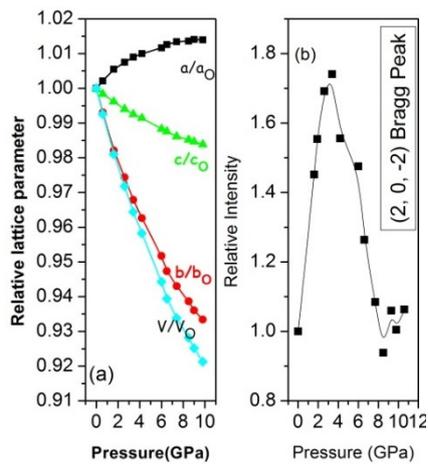

Fig.7: (a) Pressure evolution of relative lattice parameters with respect to the ambient pressure value up to 10 GPa. The monoclinic axis almost follows the unit cell volume compression indicating its importance in enhancement of magnetic frustration. (b) Relative

intensity of (2,0,-2) Bragg peak with respect to the ambient pressure value show a peak at about 3.7 GPa indicating a critical scattering behaviour arising from enhancement in antiferromagnetic order.

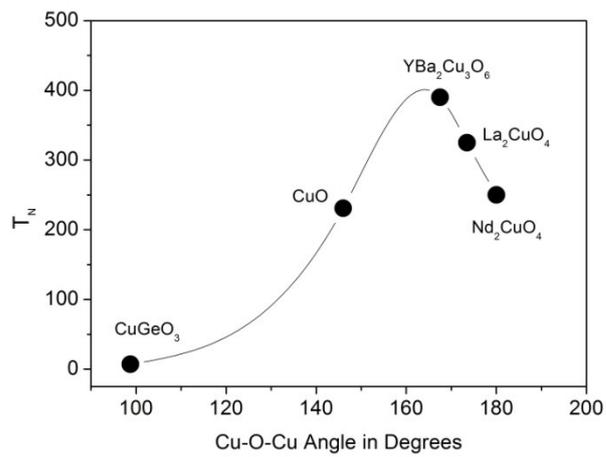

Fig.8: $T_N$ for various cuprate materials are plotted against the Cu-O-Cu bond angle. The continuous curve is a spline interpolation of the data. The references are given in the text.